# Observation of edge solitons in topological trimer arrays

Y. V. Kartashov,[1,2] A. A. Arkhipova,[1,5] S. A. Zhuravitskii,[1,3] N. N. Skryabin,[1,3] I. V. Dyakonov,[3] A. A. Kalinkin,[1,3] S. P. Kulik,[3] V. O. Kompanets,[1] S. V. Chekalin,[1] L. Torner,[2,4] and V. N. Zadkov[1,5]

[1]*Institute of Spectroscopy, Russian Academy of Sciences, 108840, Troitsk, Moscow, Russia*
[2]*ICFO-Institut de Ciencies Fotoniques, The Barcelona Institute of Science and Technology, 08860 Castelldefels (Barcelona), Spain*
[3]*Quantum Technology Centre, Faculty of Physics, M. V. Lomonosov Moscow State University, 119991, Moscow, Russia*
[4]*Universitat Politecnica de Catalunya, 08034, Barcelona, Spain*
[5]*Faculty of Physics, Higher School of Economics, 105066 Moscow, Russia*

We report the experimental observation of nonlinear light localization and edge soliton formation at the edges of fs-laser written trimer waveguide arrays, where transition from non-topological to topological phases is controlled by the spacing between neighboring trimers. We found that, in the former regime, edge solitons occur only above a considerable power threshold, whereas in the latter one they bifurcate from linear states. Edge solitons are observed in a broad power range where their propagation constant falls into one of the topological gaps of the system, while partial delocalization is observed when considerable nonlinearity drives the propagation constant into an allowed band, causing coupling with bulk modes. Our results provide direct experimental evidence of the coexistence and selective excitation in the same or in different topological gaps of two types of topological edge solitons with different internal structures, which can rarely be observed even in nontopological systems. This also constitutes the first experimental evidence of formation of topological solitons in a nonlinear system with more than one topological gap.

PhySH Subject Headings: Solitons; Topological insulators; Waveguide arrays

Topological insulators are physical structures that behave as conventional insulators in the bulk, but exhibit unusual conductive properties arising due to the existence of topologically protected, localized in-gap states at their edges that are immune to local deformations. They have been encountered in different areas of physics, including solid-state physics, acoustics, electronics, mechanics, physics of matter waves, and optics. Topological insulators have been demonstrated in different platforms and analyzed using various models, starting from dimerized Su-Schrieffer-Heeger lattices, to systems with broken time-reversal symmetry supporting unidirectional edge currents, time-periodic Floquet systems, structures with broken spatial inversion symmetry, such as valley Hall phononic or photonic crystals, various higher-order insulator platforms, and many others [1, 2]. In contrast with their widely studied electronic counterparts, photonic topological insulators [3, 4] can exhibit a considerable nonlinear response, therefore they exhibit new phenomena caused by self-action or parametric wave interactions that can alter the topological phases or affect the propagation dynamics of topological edge states. The recent progress in the investigation of passive and active nonlinear topological photonic systems is described in the reviews [5-7]. It includes, among others, nonlinearity-induced inversion of topological currents [8, 9], formation of self-sustained bulk topological currents [10, 11], nonlinear tuning of edge state energies [12], condensation and lasing in edge states in polariton systems [13-15], rich bistability effects [16, 17], enhanced parametric interactions [18-20], modulation instability of edge states [21, 22, 14], and nonlinearity-induced topological phases in systems that are topologically trivial in the linear regime [23-26].

One of the most genuine manifestations of nonlinearity in topological systems is the possibility to form edge solitons, namely self-sustained states that inherit topological protection from the linear counterparts from which they bifurcate. In multidimensional systems, such states may feature strongly asymmetric profiles reflecting their hybrid nature. In 2D settings, these solitons may form spontaneously as a result of the development of modulation instability of periodic nonlinear edge states [14, 21] (nonlinear localized modes may also form in the bulk of the insulator as demonstrated in [10, 11]). Topological soliton-like edge states that radiate in the course of propagation have been observed [27] in Floquet topological insulators [28-31], while theory for envelope quasi-solitons built on edge states has been developed for discrete [32-35] and continuous [36-39, 21] arrays of helical waveguides. Edge solitons may also form in waveguide array-based valley Hall systems [40-43]; they were observed in truncated photonic graphene lattice induced in atomic vapors [44]. Recently nonlinear corner states were demonstrated in higher-order 2D topological photonic insulators [45, 46]. Among the simplest models admitting the formation of the 1D edge solitons bifurcating from linear edge states are dimerized Su-Schrieffer-Heeger lattices [47]. Nonlinear topological states in such lattices have been studied theoretically in [48-54] and observed experimentally in electric circuits [24], topological fiber loops [55], polariton condensates [13, 56] and, in the weakly nonlinear regime, in photonic lattices [57, 58]. However, most experiments with weakly nonlinear edge states and edge solitons were performed in structures with a single topological bandgap, admitting the formation of edge solitons of only one type. The coexistence of several edge solitons with different internal structures has never been observed experimentally to date, although theoretical proposals have been recently put forward [59].

In this Letter we present the experimental observation of topological edge solitons in photonic trimer arrays, which to date had been studied only in the linear regime [60-64]. For suitable trimer separations, our system exhibits two topological bandgaps where edge states with different internal structure and symmetry exist. Edge solitons emanating from such edge states under the action of nonlinearity undergo rich bifurcations and can be traced up to the point where nonlinearity drives them into the upper allowed band, causing their coupling with bulk modes. We observed two coexisting types of edge solitons emanating from different gaps and show that they can be excited even at low powers, in contrast to conventional surface solitons in nontopological lattices that feature considerable power thresholds even in 1D settings [65-69] (the latter are substantially reduced in arrays created in photorefractive crystals [70]).

We address the propagation of light in a waveguide array consisting of $N$ trimers, which can be described by the dimensionless Schrödinger equation for the field amplitude $\psi$:

$$i\frac{\partial \psi}{\partial z} = \mathcal{H}\psi - |\psi|^2 \psi, \quad \mathcal{H} = -\frac{1}{2}\left(\frac{\partial^2}{\partial x^2} + \frac{\partial^2}{\partial y^2}\right) - \mathcal{R}(x,y), \quad (1)$$

where the function $\mathcal{R}(x,y) = p\sum_{m=1,3N} \mathcal{Q}(x-x_m, y)$ in the linear Hamiltonian $\mathcal{H}$ describes the profile of the trimer array composed from Gaussian waveguides $\mathcal{Q}(x,y) = e^{-(x^2+y^2)/a^2}$ of width $a$ and depth $p$. To perform our experiments, we fabricated arrays consisting of $N = 5$ trimers in a 10 cm-long fused silica glass sample using the fs-laser writing technique (see [71] for details). A microscope images of the fabricated arrays are presented in Fig. 1(a). The separation between waveguides inside each trimer is identical and equal to pre-selected $d = 33$ μm, at which nonlinear localization is well observed. The separation $s$ between trimers is varied in the range $18-44$ μm and controls the transition between non-topological and topological phases. Although our array is a line 1D structure, we use the 2D Eq. (1) to account for all instabilities possible in the actual continuous experimental system.

We first numerically calculate the linear spectrum of the system (all parameters as in [71]), which is central to understand the properties of the edge solitons. Linear eigenmodes of the array have the form $\psi(x,y,z) = w(x,y)e^{ibz}$, where $b$ is the eigenvalue (propagation constant) of the mode, $w$ describes the mode profile. The evolution of the linear spectrum of the array with separation $s$ between trimers is shown in Fig. 1(b) for $N = 5$. At $s \geq d$ [middle and bottom rows of Fig. 1(a)], when the inter-trimer coupling is weaker than or equal to the intra-trimer one, the system is topologically trivial. There are three bulk bands in the spectrum shown with black dots and all modes are spatially extended, see examples in Fig. 1(e) and 1(f). In contrast, at $s < d$ [top row of Fig. 1(a)] inter-trimer coupling becomes stronger than intra-trimer one, driving the system into the topological phase. This is accompanied by the appearance of *two pairs* of edge states marked by the red dots in Fig. 1(b) in each of *two* topological finite gaps in the spectrum, a remarkable distinctive feature of this system. Notice that for a sufficiently small separation $s$, a pair of modes (symmetric and antisymmetric) in a given gap is nearly degenerate [cf. modes 1 and 2 or modes 3 and 4 in Fig. 1(d)]. Modes in the top gap feature two in-phase peaks in two outermost guides (modes 1 and 2), while modes in the bottom gap feature two out-of-phase peaks in two outermost guides (modes 3 and 4). The localization of the topological edge modes progressively increases with the decrease of the value of $s$. Their appearance is consistent with the fact that the corresponding topological invariant (winding number) [3, 4]

$$\mathcal{W} = \frac{i}{2\pi}\int_{\text{BZ}} \langle w_\kappa(x,y) | \partial_\kappa | w_\kappa(x,y) \rangle d\kappa \quad (2)$$

calculated for the infinite $x$-periodic array [here $w_\kappa(x,y)$ are the normalized Bloch eigenmodes with momentum $\kappa$, the integration is carried over the first Brillouin zone of periodic array] is equal to 1, 2, and 1 for the top, middle, and bottom bands, respectively, in the topological regime at $s < d$, and is 0 for all bands in trivial regime at $s \geq d$. The spectrum for a larger array with $N = 9$ trimers [Fig. 1(c)] is practically identical – only the density of states in the bulk bands increases. The robustness of such topological states has been checked by adding small disorder into waveguide depths and positions, which did not lead to appreciable shifts of their propagation constants.

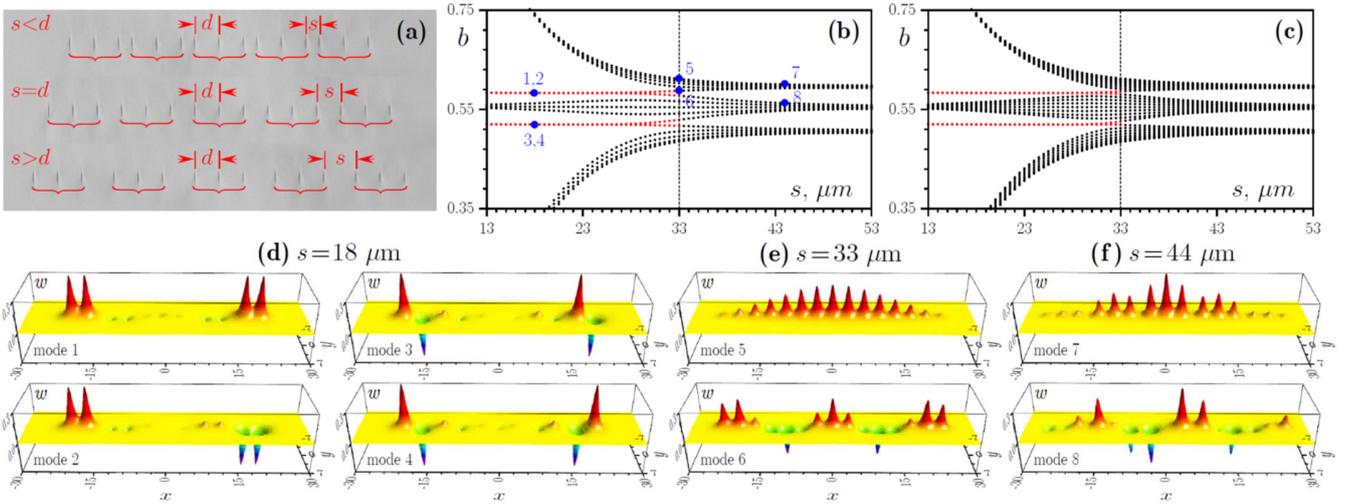

Fig. 1. (a) Microphotographs of the fs-laser written arrays of trimers in the topological and non-topological regimes. Transformation of the eigenvalue spectra upon variation of spacing $s$ for the array with $N = 5$ trimers (b) and array with $N = 9$ trimers (c). Topological branches are shown in red. Eigenmodes of the topological (d) and nontopological (e), (f) waveguide arrays correspond to the blue dots in (b).

Topological edge solitons bifurcate from linear edge states when $s < d$. They can be found in the form $\psi(x,y,z) = w(x,y)e^{ibz}$ from Eq. (1), leading to the nonlinear problem $\mathcal{H}w - w^3 + bw = 0$, where the propagation constant is now an independent variable defining the soliton amplitude and power $U = \int |\psi|^2 dxdy$. The shape of the edge soliton is determined by the symmetry of the edge state, from which such soliton bifurcates. Figure 2(a) shows the family of *out-of-phase solitons* emerging from the lower topological gap [see the top row of Fig. 2(c) for soliton shapes]. In contrast to conventional surface solitons, edge solitons do not feature an excitation power threshold and form even at low $U$ values (see state 1 whose shape may resemble that of the linear edge mode but it is a nonlinear state). When the propagation constant of such soliton shifts into the second allowed band, coupling with bulk modes occurs and the soliton acquires a long tail in the array (state 2). Several branches emerge in the vicinity of the allowed second band due to the interaction with the relevant bulk modes.

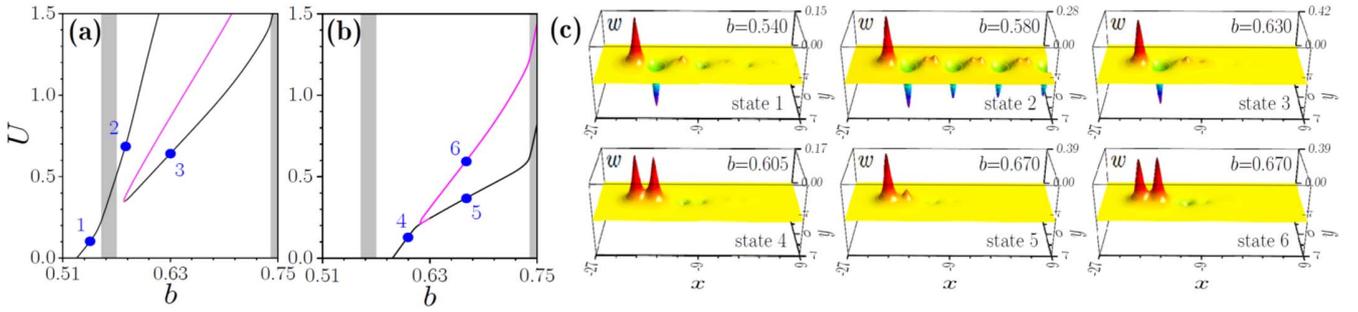

Fig. 2. Families of solitons bifurcating from out-of-phase (a) and in-phase (b) linear topological modes in array with $s=18$ μm. Black families are stable, while magenta ones are unstable. Shaded regions show bulk bands. (c) Examples of the out-of-phase (top row) and in-phase (bottom row) topological solitons corresponding to the blue dots in (a), (b).

Here we show only the simplest of them: black branches correspond to stable solitons, while magenta branches correspond to unstable ones. Soliton stability was analyzed by propagating perturbed states $\psi|_{z=0} = w(x,y)[1+\rho(x,y)]$ in Eq. (1), where $|\rho| \ll w$ is complex noise (up to 5% in amplitude), up to distances $z \sim 10^4$ that allow detecting even weak instabilities (see dynamics in [71]). The stable family with the lowest power $U$ in the top topological gap corresponds to well-localized *out-of-phase* solitons (see state 3) that remain localized until $b$ reaches the top allowed band. In turn, *in-phase* edge solitons with two in-phase left outermost peaks bifurcate directly from the linear topological edge state in the top gap [see Fig. 2(b) and state 4 in Fig. 2(c)]. This family splits into two subfamilies with increasing $b$. One of them, with the lowest $U$, is found to be stable and corresponds to a strongly asymmetric state with practically all power concentrated in the edge channel (state 5). The other family features two nearly equal in-phase peaks in the two outermost guides (state 6) and turns out to be strongly unstable. Nonlinear states localized at the edge of the array in the non-topological regime at $s \geq d$ exist only in the semi-infinite gap and feature power thresholds for their formation.

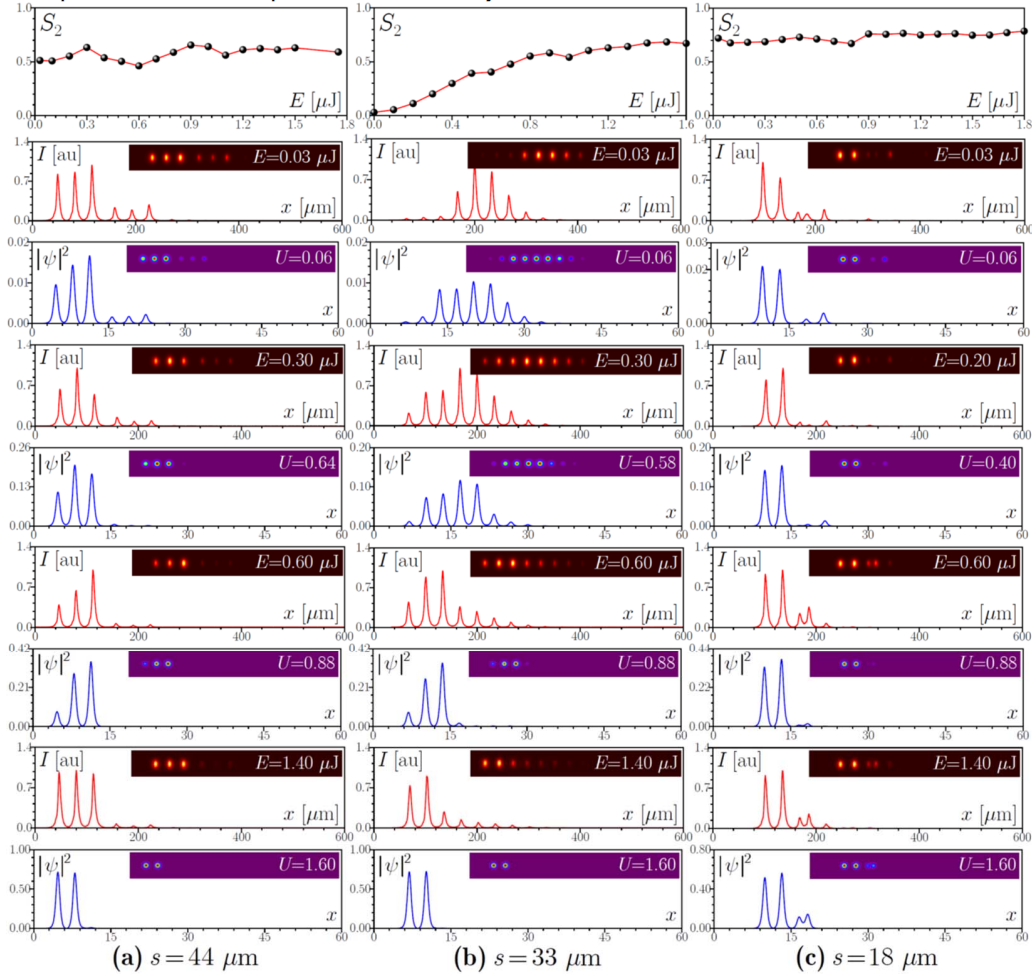

Fig. 3. Nonlinear localization and soliton formation in non-topological arrays [with $s=44$ μm (a) and $s=33$ μm (b)] and topological array [$s=18$ μm (c)] for *out-of-phase* excitation in two outermost left waveguides. The top row in each column shows the experimentally measured

fraction of energy $S_2$ concentrated in two left channels at the output versus input energy $E$ of pulses. Each column compares the experimental (red lines) and theoretical (blue lines) intensity cross-sections at $y=0$ and 2D intensity distributions (insets) for increasing $E$ values.

One of the most representative features of our system is that out-of-phase and in-phase edge solitons can coexist (in *different* or in the *same* topological gap). They exhibit qualitatively different intensity and phase structures allowing their selective excitation with properly shaped inputs. For the experimental excitation of the *out-of-phase* solitons, we use two out-of-phase beams coupled into two outermost guides and providing the largest overlap with the target state [see Fig. 2(c), states 1 and 3]. These beams are derived from a Ti:Sapphire femtosecond laser, delivering $175$ fs pulses at $1$ kHz repetition rate. To generate two independent beams, we used a Michelson interferometer with the possibility of a smooth phase change between beams. To characterize the soliton excitation efficiency we measure the fraction $S_2 = U_2/U$ of the total power remaining in two excited guides at the sample output for increasing pulse energies $E$ ($S_2$ was defined using digitized output intensity distributions from 12.3 MP scientific CMOS camera Kiralux (Thorlabs) integrated inside two circles of radius $d/2$ centered on two outermost guides, divided by total power). Typical output intensity distributions for the non-topological array with $s=44$ $\mu$m [Fig. 3(a)] show slow diffraction at low $E$ with most of the power distributing between *three* channels of the first trimer, even at the largest available energy levels. $S_2$ does not exceed 0.5-0.6 and the pattern considerably changes with $E$, so that no soliton formation is observed. In a uniform array with $s=33$ $\mu$m [Fig. 3(b)] one observes strong diffraction into the array depth in linear regime and gradual contraction of light to two excited edge guides with the increase of the pulse energy. It has to be stressed that due to the pulsed nature of the excitation the tails of spatial distributions, where the contribution from linear pulse wings may be strongest, are somewhat more pronounced in the experiment than in spatial theoretical simulations. In the uniform array, $S_2$ monotonically increases from zero to $\sim 0.7$ illustrating the existence of a power threshold for soliton formation. The picture changes qualitatively in the topological regime. Thus, for $s=18$ $\mu$m [Fig. 3(c)] both topological gaps become sufficiently wide and bands substantially narrow down [see Fig. 1(b)]. Our two-waveguide input nearly perfectly overlaps with the strongly localized out-of-phase edge soliton, thus exciting it even at the lowest power levels. A high excitation efficiency is confirmed by large values of $S_2 \sim 0.8$ practically at all energies. In this case, the second band is so narrow that when soliton crosses it, we did not observe appreciable radiation into the bulk. Such radiation is visible only for high power levels $U \sim 1.60$, when $b$ approaches the top allowed band.

To excite *in-phase* edge solitons we rely on the fact that already at moderate power levels this soliton becomes strongly asymmetric with the largest fraction of its power concentrated in the edge guide [Fig. 2(c), state 5]. Thus, we use single-spot excitation in this case. To illustrate its efficiency for $s=18$ $\mu$m, in Fig. 4 we plot the fractions of power concentrated in one, $S_1$, and two, $S_2$, outermost left guides. At lowest pulse energies $E=0.015$ $\mu$J nearly all power concentrates in the edge guide, while at $E\sim 0.235$ $\mu$J a considerable fraction of it is transferred into the second guide, as confirmed by numerical simulations. Note that this is consistent with excitation of dynamically oscillating between two guides state since the power for this input is not yet sufficient for the formation of stable asymmetric edge soliton. An asymmetric *in-phase* edge soliton forms at $E\sim 0.3$ $\mu$J and its shape remains practically unchanged over a wide energy range, again as confirmed by numerics. Notice that for this type of excitation and separation $s$, practically all light remains in the two left outermost channels of the trimer array.

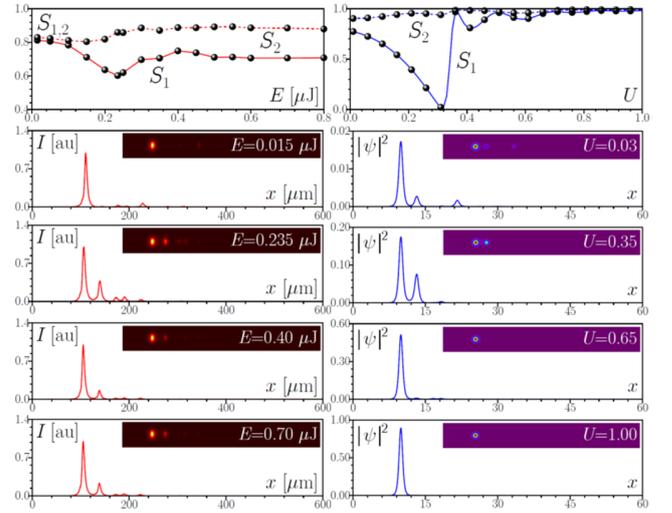

Fig. 4. In-phase edge soliton formation in the topological array with $s=18$ $\mu$m for excitation of the single left outermost waveguide (left column – experiment, right column – numerics). The top row shows fractions of energy concentrated at the output in the left outermost channel ($S_1$) or in two left channels ($S_2$) versus input energy $E$ (in experiment) or power $U$ (in numerics). Intensity cross-sections at $y=0$ and full 2D intensity distributions are shown for different input energies.

In closing, we highlight that the very nature of topological trimer waveguide arrays allows the direct exploration of different topological edge states, as they allow the realization of structures with wide topological gaps where solitons can be observed over a wide range of input powers. Our observations open the way for the investigation of nonlinear topological effects in 1D chains of dynamical or helical guides, such as Thouless pumping via edge soliton states, realization of power-controlled couplers based on topological waveguide arrays, stable lasing in edge states from different topological gaps in dissipative nonlinear systems, or the investigation of nonlinear phenomena in 2D superlattice systems with richer topological properties and spectra.

The authors acknowledge funding of this study by RSF (grant 21-12-00096). Also, support by CEX2019-000910-S [funded by MCIN/AEI/10.13039/501100011033], Fundació Cellex, Fundació Mir-Puig, and Generalitat de Catalunya (CERCA) is acknowledged.